# Finding Two Disjoint Simple Paths on Two Sets of Points is NP-Complete

Mohammadreza Razzazi[1], Abdolah Sepahvand[1,2,*]

[1] Department of Computer Engineering and Information Technology, Amirkabir University of Technology, P.O. Box 15875-4413, Tehran, Iran.

[2] School of Computer Science, Institute for Research in Fundamental Sciences (IPM), P.O .Box 19395-5746, Tehran, Iran.

E-mail: A.sepahvand@aut.ac.ir

**Abstract**

Finding two disjoint simple paths on two given sets of points is a geometric problem introduced by Jeff Erickson. This problem has various applications in computational geometry, like robot motion planning, generating polygon etc. We will present a reduction from planar Hamiltonian path to this problem, and prove that it is NP-Complete. To the best of our knowledge, no study has considered its complexity up until now. We also present a reduction from planar Hamiltonian path problem to the problem of "finding a path on given points in the presence of arbitrary obstacles" and prove that it is NP-Complete too.

Also, we present a heuristic algorithm with time complexity of $O(n^4)$ to solve this problem. The proposed algorithm first calculates the convex hull for each of the entry points and then produces two simple paths on the two entry point sets.

## 1. Introduction

This problem has various applications in path planning, VLSI etc. Assume there are two pairs of set of robots $R_1$ and $R_2$ where robots in $R_1$ and $R_2$ give set of services $s_1$ and $s_2$ respectively, sets R and B sites (points) needs set of services $s_1$ and $s_2$ respectively. The amount of time which each robot spends to give a service is not fixed. We want to find a simple path within each set of R and B so that these two paths are disjoint and the robots of one set(R or B) can be stationed at one end point of the related path to start offering the services. By choosing two simple and disjoint paths we avoid collision of robots.

In the mathematical field of graph theory, the Hamiltonian path problem and the Hamiltonian cycle problem are problems of determining whether a Hamiltonian path (a path is an undirected or directed graph that visits each vertex exactly once) or a Hamiltonian cycle exists in a given graph (whether directed or undirected). Both problems are NP-complete [1, 2].

There is a simple relation between the problems of finding a Hamiltonian path and a Hamiltonian cycle. In one point of view, the Hamiltonian path problem for graph $G$ is equivalent to the Hamiltonian cycle problem in a graph $H$ obtained from $G$ by adding a new vertex and connecting it to all vertices of $G$. Thus, finding a Hamiltonian path cannot be significantly slower (in the worst case, as a function of the number of vertices) than finding a Hamiltonian cycle. In another point of view, a graph $G$ has a Hamiltonian cycle using edge $uv$ if and only if the graph $H$ obtained from $G$ by replacing the edge by a pair of vertices of degree 1, one connected to $u$ and one connected to $v$, has a Hamiltonian path. Therefore, by trying this replacement for all edges incident to some chosen vertex of $G$, the Hamiltonian cycle problem can be solved by at most $n$ Hamiltonian path computations, where $n$ is the number of vertices in the graph [1].

The Hamiltonian cycle problem is also a special case of the travelling salesman problem, obtained by setting the distance between two cities to one if they are adjacent and two otherwise, and verifying that the total distance travelled is equal to $n$ (if so, the route is a Hamiltonian circuit; if there is no Hamiltonian circuit then the shortest route will be longer) [3].

There are several different definitions of path; simple path is a sequence of points connected to each other with line segments such that the segments do not intersect each other. Self-intersecting path is like a simple path but segments intersect each other. A close path is a simple or self-intersecting path where there exists a line segment between the first and the last point of the path.

### 1.1. **Drawing Two Disjoint Simple Paths on Two Sets of Points**

Jeff Erickson in [4] introduced the problem of finding two simple paths that have no intersection together. In technical terms, given two sets of points $X, Y$ in the plane, how can we find two disjoint simple paths from the whole points of each set or report no such paths exists. **Error! Reference source not found.** shows the problem. The points are in general position. A simple path may also be called *polygonal chain*, *polygonal curve* [5], *polygonal path* [6], *polyline* [7] or piecewise *linear curve* [7].

This problem potentially has many applications in computational geometry, for example, navigation, VLSI, robot motion planning, network design and so on.

This paper is organized as follows. Section 2 will review related studies, Section 3 will present an NP-Complete proof for the problem and reduction details, In Section 4 we propose a heuristic algorithm for problem 1.1 and in Section 5 will present the conclusion and suggestions for future works.

## 2. Related Works

Let $X$ be a finite set of points and $x_0, x_1, \ldots, x_k \in X$, be some of the points of $X$ then $L$ is called an $(s, X, t)$-path.If there exists a path which starts at $s = x_0$ and goes through vertices $x_1, \ldots, x_{k-1}$ and ends at $t = x_k$. If $\Phi$ is any subset of the plane, then we say that $L$ avoids $\Phi$ if $L$ does not intersect $\Phi$, except for possibly at points $s$ and $t$. Qi Cheng, Marek Chrobak and Gopalakrishnan Sundaram presented an NP-Complete proof for the problem of computing a simple $(s, X, t)$-path that avoids $\emptyset$.

Qi Cheng et al. in [8] also showed that the problem is solvable in polynomial time in a special case. Given a set $X$ of points inside a polygonal region $P$, and two distinguished points $s, t \in X$, they studied the problem of finding the simple polygonal paths that turn only at the points of $X$ and avoid the boundary of $P$, from $s$ to $t$. Qi Cheng et al. in [8] presented an $O(m^2 n^2)$ time and space algorithm. Xuehou Tan, Bo Jiang in [9] reviewed this problem and showed that it can be solved by $O((n^2 + m)logm)$ time, $O(n^2 + m)$ space algorithm for computing a simple path or reporting no such path exists, where $n$ is the number of points of $X$ and $m$ is the number of vertices of $P$.

Sometimes we may wish to generate a polygon that uses all points of $X$, not just a subset. This naturally leads to the problem of computing simple Hamiltonian $(s, X, t)$-paths (that is, simple $(s, X, t)$-paths that visit all points of $X$) that avoid $\Phi$. It is easy to see that the problem is NP-complete for arbitrary obstacles, so is when we restrict our attention to the case where $\Phi = P$ is a simple polygon and $X$ is inside (or outside) $P$. If $P$ is convex, a simple Hamiltonian $(s, X, t)$-path that avoids $P$ always exists and can be computed in time $O(n \, logn)$, by using angular orderings of the points in $X$ in an appropriate fashion. However, the status of this problem remains open when $P$ is an arbitrary simple polygon [8].

Alsuwaiyel and Lee [10] showed that finding a Hamiltonian $(s, X, t)$-path (not necessarily simple) in a simple polygon $P$ is NP-complete. Their proof works even in the special case when $X$ is restricted to be the vertex set of $P$. (Note that the boundary of $P$ is not a feasible solution if $s$ and $t$ are not consecutive.)

Lawrence H. Erickson and Steven M. LaValle in [11] presented a NP-Hard motion planning problem, which includes path planning in situations where crossing an obstacle is costly but not impossible, to find the path that crosses the fewest obstacles. There are, not closely related problems, including [12-14] which consider different version of finding disjoint paths between set of sources and set of targets.

## 3. Complexity Result

In this section we will prove that *drawing two disjoint simple Paths on two sets of points* (defined in Section 1.1) is NP-Complete. At first we will present the proof idea of reduction, then will prove the mentioned problem.

### 3.1. Proof Idea

A planar graph with a fixed planar embedding is called a plane graph. To prove that our problem is NP-Complete first, we prove that the problem of 'finding Hamiltonian Path in straight-line plane graph' is NP-Complete and then we reduce this special case of Hamiltonian path problem to our own problem in polynomial time.

**Theorem 1:** Finding Hamiltonian path in any (directed or undirected) planar graph is NP-Complete [15].

**Theorem 2:** Planarity testing can be conducted in linear time [16-17].

**Theorem 3:** In linear time, it is possible to find a planar embedding from a planar graph [17-19].

**Theorem 4** Any plane graph in linear time can be converted to a straight-line embedding of the graph [20-22].

We call the straight-line embedding of a plane graph as straight-line plane graph (SLPG). Algorithms for constructing planar line segment grid drawings, where the edges have integer coordinates, were developed by de Fraysseix, Pach, and Pollack [20] (shift method) and by Schnyder [21] (realizer method). They independently showed that every $n$-vertex planar graph has a planar line segment grid drawn with $O(n)$ height and $O(n)$ width, resulting in $O(n^2)$ area. . Fraysseix et al. conjectured that its complexity could be improved to $O(n)$. This bound was in fact achieved a few years later by Chrobak and Payne in [22].

**Theorem 5:** The problem of finding Hamiltonian Path in a straight-line plane graph is NP-Complete.

**Proof:** Using theorems 1, 2, 3, and 4, we simply can conclude that finding Hamiltonian path in SLPG, is NP-Complete. ∎

We then call "Hamiltonian path problem in SLPG" as *ham-path* problem.

**Lemma 1:** Any planar graph $G = (V, E)$ can be converted to two sets of points $U = V, W = \{w_1, w_2, ..., w_k\}$ ($w_i s$ are points) in the plan, such that, $\forall u, v \in V$, if $(u, v) \in E$ then, $u, v \in U$ are visible together, else $\exists w_i$ such that u, v, $w_i$ are collinear and $w_i$ is between u, v; it means, $u, v \in U$ are not visible together because $w_i$ blocks their visibility as an obstacle.

**Proof:** According to theorems 2, 3 and 4, any planar graph can be converted to a SLPG. And any SLPG $G = (V, E)$ can be converted to two sets of points $U = V, W = \{w_1, w_2, ..., w_k\}$ with the following algorithm.

| | **Convert** | **SLPG** |
|---|---|---|
| ۱ | **Input**: graph $G = (V, E)$ | |
| ۲ | **Output**: two sets of points $U, W$ with the mentioned condition in lemma 1. | |
| ۳ | **Begin** | |
| ۴ |    Set $U = V$ | |
| ۵ |    Set $k = 0$ | |
| ۶ |    Set $w = \emptyset$ | |
| ۷ |    **Define** $H = (U, E')$ a complete graph | |
| ۸ |    **For each** $(u, v) \in E'$ | |
| ۹ |    **Begin** | |
| ۱۰ |       **If** $(u, v) \notin E$ | |
| ۱۱ |       **Begin** | |
| ۱۲ |          **Define** $w_k$ a point between $u, v$ | |
| ۱۳ |          Set $W = W \cup w_k$ | |
| ۱۴ |          Set $k = k + 1$ | |
| ۱۵ |       **End** | |
| ۱۶ |    **End** | |
| ۱۷ |    **Return** $U, W$ | |
| ۱۸ | **End** | |

**Error! Reference source not found.** execution of the above algorithm on a sample input and its output (as set of red and blue points).

Clearly the condition mentioned in lemma 1 is satisfied. ∎

**Observation 1:** Consider $W$ as obstacles (blue points in **Error! Reference source not found.**); if we can find a path containing all the points in $U$ in such a way that the path does not cross the obstacles, clearly we can find a Hamiltonian path in G, because two vertices are visible in $U$, if there exists an edge between them in G.

**Theorem 6:** Finding a path on the given points and arbitrary obstacles in the plane is NP-Complete.

**Proof:** Directly concluded from lemma 1 and observation 1.

Until now we did not prove that our defined problem is NP-Complete, but in Section 3.2 we will present a reduction from *ham-path* using the above-mentioned idea. ∎

### 3.2. Details of Reduction

To prove that **"drawing two disjoint simple paths on two sets of points"** (*disjoint path* for short) problem is NP-Complete, we will reduce *ham-path* problem to it as follows.

Given a planar graph $G = (V, E)$ with line segments as edges, with the function below we can make two sets of points $U, W$ such that $U = V$, and if there exists a path containing the points of $U$, there exists a path containing the points of $W$ too.

**Definition 1:** For each vertex $v \in V$, extending each edge connected to $v$, is called an *extending edge* of $v$. These edges are called *extended edges*. This extension will divide the plane into some regions (**Error! Reference source not found.**).

| | | Convert | Function | 2 |
|---|---|---|---|---|
| ۱ | | **Input**: graph $G = (V, E)$ | | |
| ۲ | | **Output**: two sets of points with the mentioned condition in lemma 1 (and some more points as described later) | | |
| ۳ | | **Begin** | | |
| ۴ | |   Set $U = V$ | | |
| ۵ | |   Set $k = 0$ | | |
| ۶ | |   Set $w = \emptyset$ | | |
| ۷ | |   **Define** $H = (U, E')$ a complete graph | | |
| ۸ | |   **For each** $(u, v) \in E'$ | | |
| ۹ | |   **Begin** | | |
| ۱۰ | |     **If** $(u, v) \notin E$ | | |
| ۱۱ | |     **Begin** | | |
| ۱۲ | |       **Define** $w_k, w_{k+1}$ two points between $u, v$ with $\|w_k - u\| = \varepsilon$ and $\|w_{k+1} - v\| = \varepsilon$ | | |
| ۱۳ | |       Set $W = W \cup \{w_k, w_{k+1}\}$ | | |
| ۱۴ | |       Set $k = k + 2$ | | |
| ۱۵ | |     **End** | | |
| ۱۶ | |   **End** | | |
| ۱۷ | |   **For each** vertex $v \in V$ | | |
| ۱۸ | |   **Begin** | | |
| ۱۹ | |     Extend edges of $v$ | | |
| ۲۰ | |     **For each** region $r\_i$ of $v$ | | |
| ۲۱ | |     **Begin** | | |
| ۲۲ | |       **If** $\nexists w_i \in W$ in region $r\_i$ with $\|w_i - v\| = \varepsilon$ | | |
| ۲۳ | |       **Begin** | | |
| ۲۴ | |         **Define** $w_k$ a point in region $r\_i$ with $\|w_k - v\| = \varepsilon$ | | |
| ۲۵ | |         Set $W = W \cup w_k$ | | |
| ۲۶ | |         Set $k = k + 1$ | | |
| ۲۷ | |       **End** | | |
| ۲۸ | |     **End** | | |
| ۲۹ | |     **For each** *extended edge* $e$ of $v$ | | |
| ۳۰ | |     **Begin** | | |
| ۳۱ | |       **Define** $w_k$ a point on $e$ with $\|w_k - v\| = \varepsilon$ | | |
| ۳۲ | |       Set $W = W \cup w_k$ | | |
| ۳۳ | |       Set $k = k + 1$ | | |
| ۳۴ | |     **End** | | |
| ۳۵ | |   **End** | | |
| ۳۶ | |   **Return** $U, W$ | | |
| ۳۷ | | **End** | | |

The output of **Convert Function 2** includes two sets of points $U, W$, such that $U = V$ and $W$ contains some points that satisfy the condition of lemma 1 and some other points. Points inserted into $W$ are shown in **Error! Reference source not found.**. These points are added in such a way that all of them can be connected to each other as a cycle if there exists a path containing the points of $U$.

**Claim 1:** If there exists a path containing the points of $U$, there exists a cycle containing the points of W.

**Proof:** Points added to $W$ are added in such a way that to guarantee the above claim. If there exists a simple path containing $U$, we added the points to $W$ such that we could have a cycle just moving near the path with about epsilon distance from it (**Error! Reference source not found.**).

To explore more, let $L = \{p_1, p_2, ..., p_k\}$ be the path that contains the points of $U$, such that $p_1, p_k$ are the end points (just one edge is connected to them). In the above algorithm we inserted at least three points to $W$ with epsilon distance from these end points (**Error! Reference source not found.** (a)).

Knowing this, we can connect these points (three or more) together as is shown in **Error! Reference source not found.** (b) and **Error! Reference source not found.**. We need these properties to make a simple path containing the points of W.

For $v = p_i, 1 < i < k$, there exists at least five points in $W$ that are on the concave side of point $p_i$ with distance epsilon from $p_i$ (**Error! Reference source not found.** (a)), and there exists at least one point in $W$ that is on the convex side of the point $p_i$ with distance epsilon from $p_i$ (**Error! Reference source not found.** (a)).

We can easily connect all such points on the concave side as shown in **Error! Reference source not found.** (b). This trick is useful for turning around $p_i, 1 < i < k$ and to build a simple path containing all the points of W. By continuing these connecting points (as mentioned), there will be a chain containing the points of W. ∎

As mentioned in observation 1 and claim 1, we can find a path from $U$, if and only if, we can find a Hamiltonian path from $V$. Hence we have the following theorem.

**Theorem 7:** "Drawing two disjoint simple paths on two sets of points" is NP-Complete.

**Proof:** clearly if we have two disjoint simple paths, simply we can verify their disjointness in polynomial time, then the problem is NP. Because of the reduction mentioned in this section, we can conclude that the problem is NP-Complete. ∎

## 4. The Proposed Algorithm

In this section, we present a heuristic algorithm in order to find two simple and distinct Paths from the set of entry points, namely red ($R$) and blue ($B$). The objective is to minimize the intersection points between the two obtained Paths.

First, we separately generate the convex hull for each of $R$ and $B$ sets and we call the set of edges of these convex hulls $RCH$ and $BCH$, respectively (Figure 8 (b)). By randomly removing an edge from RCH and BCH so that it has the highest number of intersections with other $CH$, we obtain two simple red and blue paths (Figure 8 (c)). Then, according to **Algorithm 1**, we add to each path, depending on the color of the path, those points in $R$ or $B$ sets that are not members of the path. Assume $p$ is any given point and $q$ is one of the two end points of a path $x$, we say $q$ is visible from $p$ (similarly $p$ is visible from $q$) if $pq$ ($qp$) does not intersect path $x$, except at $q$ ($p$). The point $p$ is visible form edge $(u, v)$ of path $x$, if neither edge $(p, u)$ nor $(p, v)$ intersect with path $x$ (also edge $(u, v)$ is visible from point $p$). For example, in Figure 8 (d), point $rp$ is visible from $e_1, e_4, p_1$ and $p_2$ of the red path, and point $rp$ is not visible form $e_2$ and $e_3$ of the red path.

| Algorithm 1 |
|---|
| **Input:** two sets of red ($R$) and blue ($B$) points. |
| **Output**: Two simple path from each set. |
| ۱    **Generate** the convex hull for each of $R$ and $B$ sets (Figure 8.b). We call the set of edges of these convex hulls $RCH$ and $BCH$, respectively. |
| ۲ |
| ۳    **Let** $V(RCH)$ and $V(BCH)$ be set of endpoints of edges in $RCH$ and $BCH$ respectively. |
| ۴    **Initialize** the set $M = \emptyset, N = \emptyset, V = \emptyset \text{ and } W = \emptyset$. |
| ۵    **Add** edges from $RCH$ to $M$ such that they have the most crossing points with $BCH$. |
| ۶    **Add** edges from $BCH$ to $N$ such that they have the most crossing points with R$CH$. |
| ۷    **Choose** edge $(u_r, v_r) \in M$ randomly and remove $(u_r, v_r)$ from $RCH$. |

| | |
|---|---|
| 8 | **Choose** edge $(u_b, v_b) \in N$ randomly and remove $(u_b, v_b)$ from $BCH$. |
| 9 | **Do** { |
| 10 |    **If** $(R - V(RCH) \neq \emptyset)$ |
| 11 |    { |
| 12 |      **Choose** point $rp \in R - V(RCH)$ randomly. |
| 13 |      Add $u_r, v_r$ and all edges from $RCH$ to $W$ where they are visible from $rp$. |
| 14 |      **If** $u_r, v_r$ and all edges from $RCH$ are not visible from $rp$ and there exist unprocessed points in |
| 15 |      $R - V(RCH)$ |
| 16 |        Go to step 12 and select another point. |
| 17 |      **Else** |
| 18 |        Restart the program. |
| 19 |      **Choose** $w \in W$ randomly. (* it can be either a point or an edge*). |
| 20 |      **If** ( $w \in \{u_r, v_r\}$ ) { |
| 21 |        Connect $rp$ to $w$ and add edge $(rp, w)$ to $RCH$. |
| 22 |        **If** $(w = u_r)$ |
| 23 |          $u_r = rp$ (* $u_r$ is now the new endpoint of the red path*). |
| 24 |        **Else** |
| 25 |          $v_r = rp$ (* $v_r$ is now the new endpoint of the red path*). |
| 26 |      } **Else** (* an edge is chosen*) |
| 27 |        Remove $w = (w_1, w_2)$ from $RCH$ and insert $(w_1, rp)$ and $(w_2, rp)$ to $RCH$. |
| 28 |      Set $W = \emptyset$. |
| 29 |    } |
| 30 |    **If** $(B - V(BCH) \neq \emptyset)$ |
| 31 |    { |
| 32 |      **Choose** point $bp \in B - V(BCH)$ randomly. |
| 33 |      Add $u_b, v_b$ and all edges from $BCH$ to V where they are visible from b$p$. |
| 34 |      **If** $u_b, v_b$ and all edges from $BCH$ are not visible from $bp$ and there exist unprocessed points in |
| 35 |      $B - V(BCH)$ |
| 36 |        Go to step 32 and select another point. |
| 37 |      **Else** |
| 38 |        Restart the program. |
| 39 |      **Choose** $v \in V$ randomly. (* it can be either a point or an edge*). |
| 40 | |
| 41 |      **If** ( $v \in \{u_b, v_b\}$ ) { |
| 42 |        Connect $bp$ to $v$ and add edge $(bp, v)$ to $BCH$. |
| 43 |        **If** $(v = u_b)$ |
| 44 |          $u_b = bp$ (* $u_b$ is now the new endpoint of the blue path*). |
| 45 |        **Else** |
| 46 |          $v_b = bp$ (* $v_b$ is now the new endpoint of the blue path*). |
| 47 |      } **Else** (* an edge is chosen*) |
| 48 |        Remove $v = (v_1, v_2)$ from $BCH$ and insert $(v_1, bp)$ and $(v_2, bp)$ to $BCH$. |
| 49 |      Set $V = \emptyset$. |
| 50 |    } |
| 51 | } **While** $(R - RCH \neq \emptyset$ or $B - BCH \neq \emptyset)$. |

In the above pseudo code, we first generate the convex hull of two sets. Then, we remove one of the edges which has the most intersection with other convex hull randomly (5-8). We start to add free[1] red and blue points to corresponding chains (9-51). Lines 10-29 are adding free red point to the red chain and lines 30-50 are adding free blue points to blue chain. A red point $rp$ is added to a red path $x$, by doing the following:

$rp$ is connected to any endpoint of the path $x$ which is visible from $rp$, otherwise, it finds an edge $(u,v)$ of path $x$ which is visible from $rp$, removes $(u,v)$ from path $x$, and adds edges $(rp,u)$ and $(rp,v)$ to path $x$. Similar steps are taken for adding a blue point to the blue path. In rare cases that such an endpoint or edge cannot be found (see section 4.2), algorithm is restarted. For example Figure 8 is showing these operations on given sample red and blue points.

### 4.1. The Proposed Algorithm Analysis

Assuming that $n = max${the number of blue and red points}, calculating the convex hull using Graham's algorithm takes $O(nlogn)$. Checking whether the two edges have any intersection with each other or not takes a constant time. Steps 5-8 takes $O(n^2)$ time at most. Most of the time needed for adding a point $p$ to a path belong to finding the visible edges of a path from the point $p$, this takes $O(n^3)$ time. We add $n$ points to the path, so that the time complexity of the entire algorithm becomes $O(n^4)$. In finding the above complexity, we used naive algorithms. Clearly, there are more efficient algorithm for visibility and finding intersections of two convex hulls, which by using them our complexity could be reduce significantly.

### 4.2. A Special Case of the Algorithm

A condition may occur such that none of the remaining points from a point set is visible from any of the edges or points at the two ends of the path. If this special case occurs, we execute Algorithm 1 from the beginning. Figure 9 shows an example of this special case.

We executed the proposed algorithm 100,000 times on a point set which included the set point of Figure 9, and in 99,742 executions the algorithm successfully produced a simple path in the first run. Also, in order to successfully test the proposed algorithm, we used point sets cardinalities: 10, 20, 30, …, 100, 130, 160, 190, and 220. For each cardinality, we randomly generated 1000 pair of point sets, one for set of red points and the other for set of blue points. The proposed algorithm was executed 10,000 times on each pair of point sets. Figure 10 shows the probability of algorithm's success at the first run.

### 4.3. The Proposed Algorithm Test

In order to test the proposed algorithm, we used point sets cardinalities: 10, 20, 30, …, 90, 100. For each cardinality, we randomly generated 100 pairs of point sets, one for set of red points and the other for set of blue points. We executed the proposed algorithm 1000 times on each pair of point sets. Table 1 shows the obtained results.

---

[1] We call a red (blue) point free if it is not connected to the red (blue) chain yet.

We designed an exact algorithm, with exponential time complexity, to obtain the optimum solution. Using this exact algorithm, we carried out the following tests: we used point set cardinalities: 5,6,7, …, 11, 12 for each cardinality, we randomly generated 100 pairs of point sets, each pair consisting of a set of red points and a set of blue points, with both sets having the same cardinality. For each pair of point sets, we ran the proposed algorithm 1000 times and calculated the average and the maximum number of intersections. Table 2 shows the results.

## 5. Conclusion and Future Works

In this paper we presented proof of NP-Completeness for finding two disjoint simple paths on two given sets of points. Also we proved that finding a path on a given set of points in presence of arbitrary obstacles is NP-Complete. These proofs are done by reduction from planar Hamiltonian path problem. Finding two disjoint paths on two given sets of points has application in robots motion planning, polygon generation etc. Also, we proposed a heuristic algorithm to solve this problem in polynomial time, the objective of which was to minimize the number of intersection points between the two paths.

We discussed the problem in two dimensional space. As a future work this problem can be generalized to higher dimensions. Another interesting problem is finding a path on the given points while a path as an obstacle exists. By solving this problem, some problems mentioned in [8] may be easily solved.

## 6. Acknowledgment

The authors would like to thank the anonymous reviewers for their valuable comments and suggestions to improve the quality of the paper. This research was in part supported by a grant from IPM. (No. CS1393-4-47).

**Biographies**

**Mohammadreza Razzazi** received the M.S. degree in Computer Science from Stanford University and the Ph.D. degree in Computer Science from the University of California at Santa Barbara. Currently he is an associate professor in computer engineering & IT department at Amirkabir University of Technology (AUT) at Tehran. His primary areas of research are computational geometry, robotics, and computer graphics.

**Abdolah Sepahvand** received the BSc degree in Computer Engineering in 2013 from Shahid Rajaee Teacher Training University (SRTTU) and MSc degree in Computer Engineering in 2015 from Amirkabir University of Technology (AUT). His research interests include computational geometry and approximation algorithms.

**Figure and table captions**

Figure 1: Given two set of red and blue points, find a path from red points (L1), and a path from blue points (L2), such that L1 and L2 are disjoint [7].

Figure 2: (a) Input graph $SLPG$ $G = (V, E)$; (b) Complete graph $H = (U, E')$. Blue edges are not in G; (c) Output of the algorithm that includes two sets of points with the conditions mentioned in lemma 1.

Figure 3: Extending edges of vertex $v$ and respective regions. Dash lines are extended edges.

Figure 4: In (a) red points and red lines represent input graph $G = (V, E)$ and all line segments (blue and red) represent the complete graph $H = (U, E')$. (b) Lines 8-16 will add blue points to $W$. (c) Black lines are extended edges of $v \in V$ (line 19). (d) Lines 19-28 will add green points to $W$. (e) Lines 29-34 will add purple points to $W$. (f) Output of the algorithm, red points are in $U$ and other points are in $W$ (different color for points are used for well understanding and all points in $W$ play the same role).

Figure 5: Red line segments are a path containing the points of $U$. Black line segments are a cycle containing the points of $W$.

Figure 6: (a) There are at least three points in $W$ with distance epsilon from the end point $v$. (b) We can easily connect points with epsilon distance from $v$.

Figure 7: There exist at least five points on the concave side of vertex v and exist at least one point on its convex side.

Figure 8: (a) Represents the entry points sets. (b) Convex hull of red and blue points sets are generated and the $re$ edge is selected randomly from the convex hull of red points set and the $be$ edge from the convex hull of blue points set in order to be removed. (c) Omission of $re$ and $be$ edges from convex hull. (d) The $rp$ point is selected to be added to the red path and is $visible$ from $e_1, e_4, p_2$ $and$ $p_1$. (e) Represents the output of applying the algorithm.

Figure 9. The point $p$ is not visible from any of the edges or end points of a path

Figure 10. The probability of success of the proposed algorithm at the first run on the set of entry points

Table 1. The results obtained from the implementation of the proposed algorithm.

Table 2. The results of executing the optimal algorithm and the proposed algorithm.

**FIGURE LEGENDS**

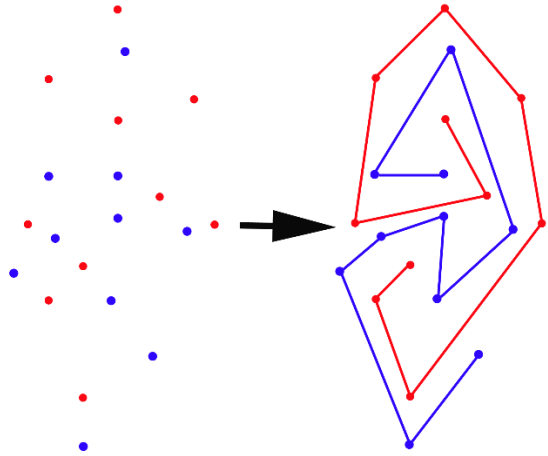

Figure 1: Given two set of red and blue points, find a path from red points (L1), and a path from blue points (L2), such that L1 and L2 are disjoint [7].

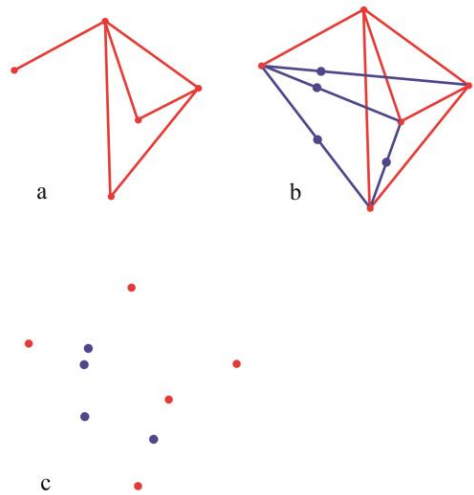

Figure 2: (a) Input graph $SLPG$ $G = (V, E)$; (b) Complete graph $H = (U, E')$. Blue edges are not in G; (c) Output of the algorithm that includes two sets of points with the conditions mentioned in lemma 1.

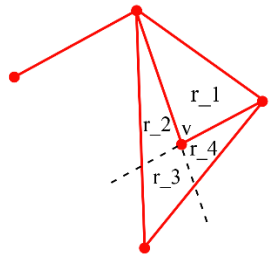

Figure 3: Extending edges of vertex $v$ and respective regions. Dash lines are extended edges.

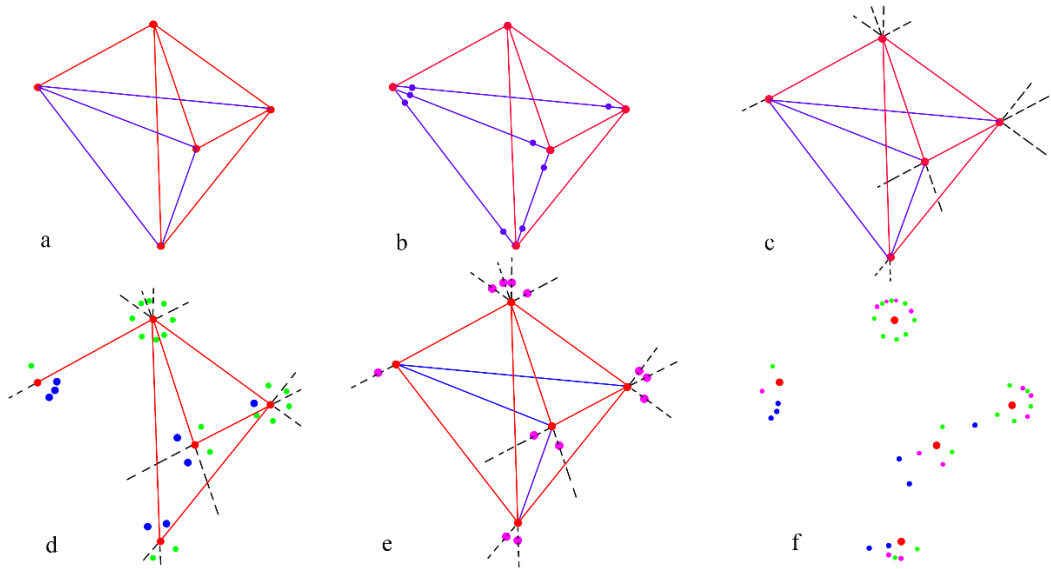

Figure 4: In (a) red points and red lines represent input graph $G = (V, E)$ and all line segments (blue and red) represent the complete graph $H = (U, E')$. (b) Lines 8-16 will add blue points to $W$. (c) Black lines are extended edges of $v \in V$ (line 19). (d) Lines 19-28 will add green points to $W$. (e) Lines 29-34 will add purple points to $W$. (f) Output of the algorithm, red points are in $U$ and other points are in $W$ (different color for points are used for well understanding and all points in $W$ play the same role).

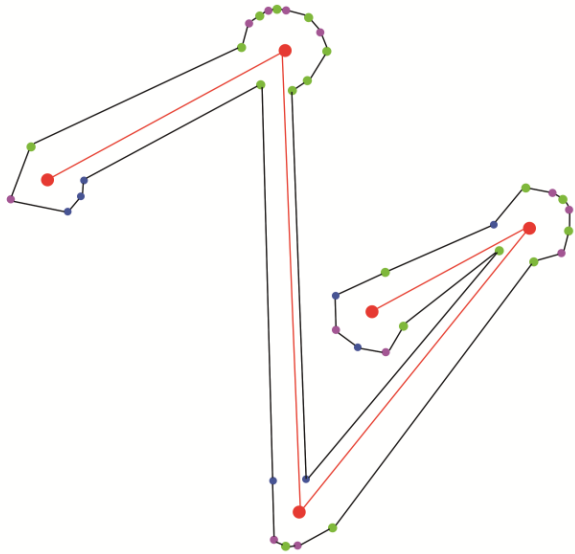

Figure 5: Red line segments are a path containing the points of $U$. Black line segments are a cycle containing the points of $W$.

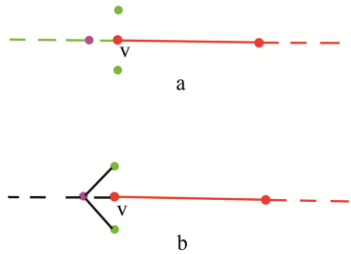

Figure 6: (a) There are at least three points in $W$ with distance epsilon from the end point $v$. (b) We can easily connect points with epsilon distance from $v$.

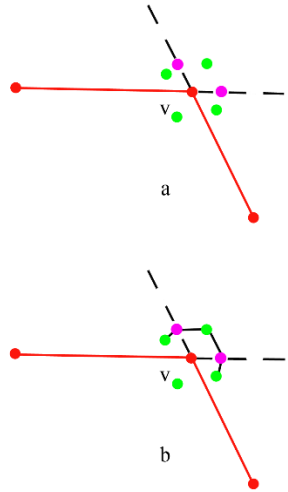

Figure 7: There exist at least five points on the concave side of vertex v and exist at least one point on its convex side.

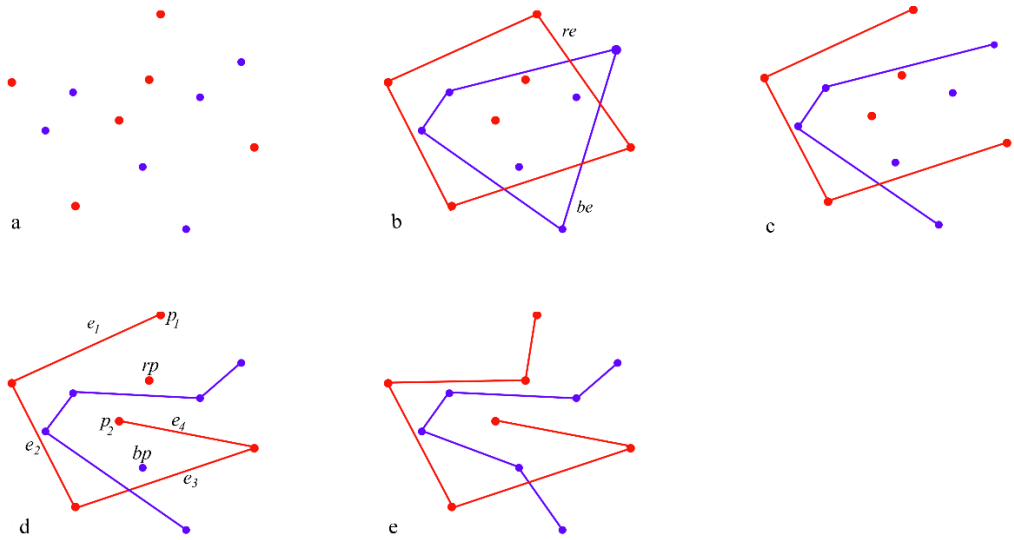

Figure 8: (a) Represents the entry points sets. (b) Convex hull of red and blue points sets are generated and the $re$ edge is selected randomly from the convex hull of red points set and the $be$ edge from the convex hull of blue points set in order to be removed. (c) Omission of $re$ and $be$ edges from convex hull. (d) The $rp$ point is selected to be added to the red path and is $visible$ from $e_1, e_4, p_2$ $and$ $p_1$. (e) Represents the output of applying the algorithm.

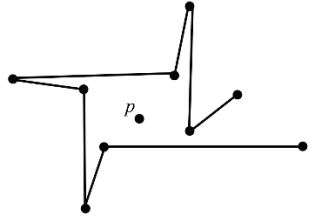

Figure 9. The point $p$ is not visible from any of the edges or end points of a path

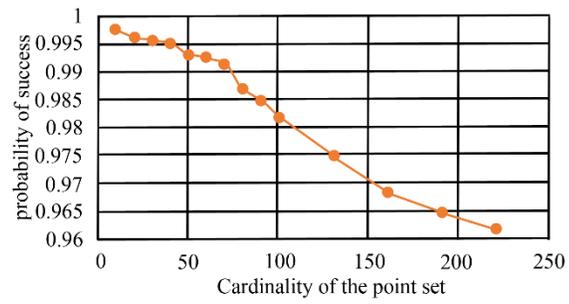

Figure 10. The probability of success of the proposed algorithm at the first run on the set of entry points

**TABLES**

Table 1. The results obtained from the implementation of the proposed algorithm.

| Number of entry points from each color | The minimum number of created intersections by the proposed algorithm | The maximum number of created intersections by the proposed algorithm | The average number of created intersections by the proposed algorithm |
|---|---|---|---|
| 10 | 0 | 7 | 1.95 |
| 20 | 0 | 19 | 5.41 |
| 30 | 1 | 27 | 10.63 |
| 40 | 1 | 35 | 15.16 |
| 50 | 4 | 47 | 19.10 |
| 60 | 8 | 53 | 24.87 |
| 70 | 11 | 60 | 29.59 |
| 80 | 16 | 77 | 33.77 |
| 90 | 17 | 81 | 37.63 |
| 100 | 22 | 86 | 37.54 |

Table 2. The results of executing the optimal algorithm and the proposed algorithm.

| Number of entry points from each color | The maximum number of intersections of the proposed algorithm | The average number of intersections of the optimal solution | The average number of intersections of the proposed algorithm |
|---|---|---|---|
| 5 | 3 | 0.34 | 0.64 |
| 6 | 5 | 0.63 | 0.83 |
| 7 | 4 | 0.91 | 1.17 |
| 8 | 6 | 0.98 | 1.34 |
| 9 | 5 | 1.11 | 1.5 |
| 10 | 5 | 1.13 | 1.91 |
| 11 | 6 | 1.43 | 2.13 |
| 12 | 9 | 1.57 | 2.51 |